\documentclass[aip,amsmath,amssymb,reprint]{revtex4-1}

\usepackage{graphicx}
\usepackage[colorlinks=true,citecolor=blue,breaklinks=true,linkcolor=blue,linktocpage=true,urlcolor=blue,pagebackref=false]{hyperref}
\usepackage[utf8]{inputenc}
\usepackage[T1]{fontenc}
\usepackage{orcidlink}
\usepackage{mathptmx}

\newcommand{\llangle}{\langle\!\langle}
\newcommand{\rrangle}{\rangle\!\rangle}

\newcommand{\void}[1]{}

\makeatletter
\def\@email#1#2{%
 \endgroup
 \patchcmd{\titleblock@produce}
  {\frontmatter@RRAPformat}
  {\frontmatter@RRAPformat{\produce@RRAP{*#1\href{mailto:#2}{#2}}}\frontmatter@RRAPformat}
  {}{}
}%
\makeatother

\begin{document}

\title{The open driven two-level system at conical intersections of quasienergies}
\author{Sigmund Kohler\,\orcidlink{0000-0003-3668-8030}}
\email{sigmund.kohler@csic.es}
\affiliation{Instituto de Ciencia de Materiales de Madrid, CSIC, E-28049 Madrid, Spain}
\date{\today}

\begin{abstract}
We study the stationary state of an ac-driven two-level system under
particle exchange with a fermionic environment.  A particular
question addressed is whether there exist limits in which the
populations of the Floquet states are determined by their quasienergies or
their mean energies, respectively.  The focus lies on parameters in the
vicinity of conical intersections of quasienergies, because there the two
kinds of energies behave rather differently, such that the characteristics
of the two intuitive limits are most pronounced.  A main finding is a
crossover from a Floquet-Gibbs-like state at low temperatures to a mean-energy
dominated state at intermediate temperatures.  Analytical estimates are
confirmed by numerical calculations.
\end{abstract}

\maketitle

\section{Introduction}

Quantum systems at thermal equilibrium are usually assumed to be in a Gibbs
state which, depending on whether one allows for particle exchange, is
the canonical or the grand canonical ensemble.  For a more profound
description, one may employ system-bath models that allow for the exchange
of energy \cite{Weiss2012} or particles \cite{BruusFlensberg2004} with
an environment to obtain the dynamics and the stationary solution of the
reduced system density operator.  Usually at least for very weak coupling,
the latter eventually becomes a Gibbs state.\cite{LandiRMP22}  For driven
systems, the situation is more involved, because no such formal expression
for the stationary density operator is known.  Hence, for its computation,
system-bath models are indispensable.  Notably, then even in the
weak-coupling limit, the stationary state may qualitatively depend on
details of the system-bath interaction.\cite{FerronPRL12, BlattmannPRA15}
Moreover, genuine non-equilibrium effects emerge such as pumping and
rectification of heat \cite{SegalPRL05, SegalPRE06} and charge
\cite{LehmannPRL02, LehmannJCP03} currents.  These non-equilibrium effects
are particularly important for Floquet engineering which is the design of
effective static Hamiltonians by ac fields.\cite{GrossmannPRL91,
GrossmannEL92, HolthausPRL92, EckardtRMP17}
In exceptional cases, driven dissipative systems assume
Floquet-Gibbs states, i.e., canonical states in which the eigenenergies are
formally replaced by quasienergies.\cite{ThingnaJCP12, ShiraiPRE15,
ShiraiNJP16, IwahoriPRB16, MoriARCMP23}  There are also situations in
which, by contrast, the mean energies of the Floquet states determine the
populations.\cite{KohlerPRE98, KetzmerickPRE10, KohlerPRA24}

In practice none of these limits will be perfectly realized.  Nevertheless
one may observe a clear tendency towards the one or the other.  To explore this
behavior, the vicinity of conical intersections of quasienergies turned out
to be rather interesting.\cite{KohlerPRA24}  Such intersections emerge when a
spatio-temporal symmetry of a driven quantum system such as generalized
parity \cite{PeresPRL91} allows the exact crossing of quasienergies.  A weak
perturbation may break this symmetry, so that the former crossing turns
into an accidental degeneracy in a two-dimensional parameter space.  A most
interesting feature of canonical intersections in driven systems is that
along lines with constant quasienergies, the mean energies of the Floquet
states are interchanged.\cite{KohlerPRA24}  Therefore in this regime, the
roles played by these energies will be most pronounced.  For example, there
may emerge discontinuities of the populations \cite{EngelhardtPRL19} or a
crossover from a mean-energy state to a more exceptional Floquet-Gibbs
state.\cite{KohlerPRA24}

In this work, we explore how the results of Ref.~\onlinecite{KohlerPRA24}
for the driven dissipative two-level system are affected by replacing the
bosonic heat bath with electron reservoirs.  To stay close to that situation,
electron-electron interaction and the spin degree of freedom are ignored.
A suitable description of such systems is a Floquet-based master equation
for the single-particle density matrix.  It was derived in the context
of molecular wires to study ratchet effects \cite{LehmannPRL02} and current
rectification by ac fields. \cite{LehmannJCP03}

This work is structured as follows.  In Sec.~\ref{sec:model}, we introduce
conical intersections of quasienergies as a consequence of a
spatio-temporal symmetry, the coupling to a fermionic particle reservoir,
and a master equation formalism for the single-particle density operator
based on Floquet theory.  Section \ref{sec:results} is devoted to
approximative solutions of the master equation which lead to conjectures
for the stationary state which are confirmed by numerical calculations.
Conclusions are drawn in Sec.~\ref{sec:conclusions}, while the properties
of Floquet states close to conical intersections are summarized in the
appendix.

\section{Open driven two-level system}
\label{sec:model}

\subsection{Driven two-level system and conical intersections}
\label{sec:cone}

\begin{figure}[b]
\centerline{\includegraphics{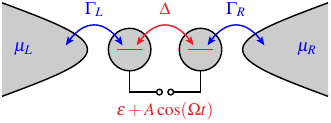}}
\caption{Driven two-level system with tunnel coupling $\Delta$ and
ac-modulated onsite energies.  The wire-lead tunnel couplings
$\Gamma_{L,R}$ enable particle exchange with the respective site.}
\label{fig:setup}
\end{figure}

\begin{figure*}
\centerline{\includegraphics{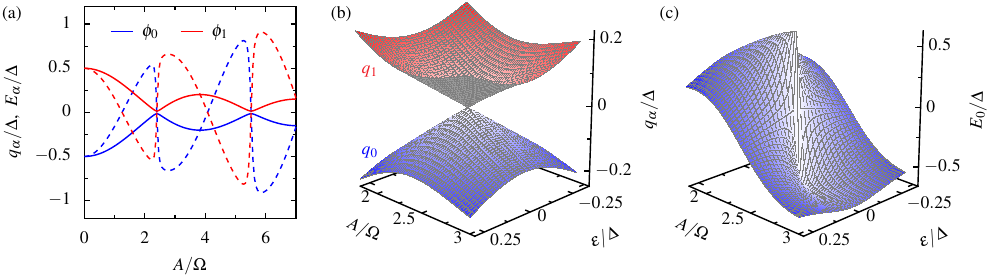}}
\caption{(a) Quasienergies (solid lines) and mean energies (dashed) for
driving frequency $\Omega=10\Delta$ and the rather small detuning
$\varepsilon=0.03\Delta$ as function of the driving amplitude.  For
$\varepsilon=0$, owing to generalized parity the avoided crossing would be
exact.  (b) Quasienergies as a function of the detuning and the driving
amplitude in the vicinity of the first quasienergy crossing observed in
panel (a) close to $\varepsilon=0$ and $A_0\approx 2.4\Omega$.  (c) Mean
energy of the Floquet state $|\phi_0(t)\rangle$ with---for the Brillouin
zone centered at zero---lower quasienergy.}
\label{fig:cone}
\end{figure*}

We consider a two-level system with tunable onsite energies and tunnel
coupling to two leads as is sketched in Fig.~\ref{fig:setup}.  When
occupied with a single electron, the Hamiltonian of the central system in
pseudo-spin notation reads
\begin{equation}
H(t) = \frac{\Delta}{2}\sigma_x
      + \frac{1}{2}[\varepsilon + A\cos(\Omega t)]\sigma_z \,,
\label{H1p}
\end{equation}
where $\sigma_{x,z}$ are the usual Pauli matrices.  The two sites are
tunnel coupled with strength $\Delta$ and possess a static detuning
$\varepsilon$.  The ac driving enters via a sinusoidal additional detuning
with amplitude $A$ and frequency $\Omega$.  We use units with $\hbar=1$.

A suitable tool for treating periodically time-dependent quantum systems is
Floquet theory.  Its cornerstone is discrete time translation invariance
from which follows that the Schr\"odinger equation has a complete set of
solutions of the form $e^{-iqt}|\phi(t)\rangle$ with quasienergy $q$ and
Floquet state $|\phi(t)\rangle = |\phi(t+2\pi/\Omega)\rangle$.
\cite{ShirleyPR65, SambePRA73, Hanggi98} The Floquet states and their
quasienergies are eigensolutions of the operator $\mathcal{H} = H(t)
-i\partial/\partial t$ in Sambe space, which is the direct product of
Hilbert space and the space of $2\pi/\Omega$-periodic
functions.\cite{SambePRA73}  Quasienergies correspond to the phase that a
Floquet state acquires during one driving period.  An important property of
a Floquet state is its mean energy $E$, i.e, its energy expectation value
averaged over one driving period.  It relates to the quasienergy via a
geometric phase,\cite{AharonovPRL87, MooreJPA90b}
\begin{equation}
E = q + \llangle\phi(t)|i\partial_t|\phi(t)\rrangle
= q + \sum_k k\Omega \langle\phi_k|\phi_k\rangle ,
\label{idt}
\end{equation}
where the outer angles denote time average, while $|\phi_k\rangle$ is the
$k$th Fourier coefficient of the Floquet state $|\phi(t)\rangle$.

When $|\phi(t)\rangle$ is a Floquet state with quasienergy $q$, then for
any integer $n$, $e^{-in\Omega t}|\phi(t)\rangle$ is a physically
equivalent Floquet state with quasienergy $q+n\Omega$.  All equivalent
states correspond to the same solution of the Schr\"odinger equation.
Therefore, it is sufficient to consider a particular one.  Nevertheless in
the present context, there exists a unique meaningful choice, namely the
one for which upon adiabatically reducing the driving amplitude to zero,
$q$ and $E$ become equal.  Only then we can expect the emergence of a
meaningful Floquet-Gibbs state.  This choice can be ensured by considering
large driving frequencies and choosing the Brillouin zone such that it
comprises the spectrum of the undriven Hamiltonian.\cite{ShiraiNJP16}  This
implicitly fixes the ordering of the Floquet states such that we can
attribute the ``ground state index $0$'' to the state with lower
quasienergy.

For $\varepsilon=0$, both the Hamiltonian \eqref{H1p} and the Floquet
Hamiltonian $\mathcal{H}$ possess a spatio-temporal $\mathbb{Z}_2$
symmetry, namely the generalized parity $G = \sigma_x P$, where
$P=e^{(\pi/\Omega)\partial_t}$ shifts time by half a driving
period.\cite{PeresPRL91}  For $2\pi/\Omega$-periodic functions, $G^{-1}=G$
which implies that the eigenvalues of $G$ are $\pm1$.  Notably the operators
$i\partial_t$, $\sigma_x$, and $\sigma_z\cos(\Omega t)$ are invariant under
transformation with $G$ such that $[G,\mathcal{H}] = 0$.  Therefore, the
Floquet states are also eigenstates of $G$ or, in the case of degeneracies,
can be chosen as such.  Moreover, they can be classified as even or odd,
depending on the respective eigenvalue of $G$.  It is known that
quasienergies from different symmetry classes may form exact
crossings.\cite{Haake2018}

Since $G\sigma_z G^{-1} = -\sigma_z$, a non-vanishing detuning $\varepsilon$
breaks the generalized parity of $\mathcal{H}$.  As a consequence, exact
quasienergy crossings become avoided, as can be witnessed in
Fig.~\ref{fig:cone}(a).  At avoided crossings, the associated states and,
thus, their expectation values are interchanged.  This is reflected by the
behavior of the mean energies which form exact crossings when quasienergies
anti-cross.  Moreover, in the parameter space of detuning and
driving amplitude, $(\varepsilon,A)$, one finds a degeneracy only at
isolated points, such that the quasienergies form conical intersections,
see Fig.~\ref{fig:cone}(b).  The behavior of the mean energy of the state
on the lower cone is sketched in Fig.~\ref{fig:cone}(c). It visualizes the
continuous exchange of the mean energies when going around the cone tip
along a line with constant quasienergy.\cite{KohlerPRA24} Accordingly,
whenever the mean energies govern dissipation, we can expect a significant
parameter dependence of the stationary state.  In turn, for a Floquet-Gibbs
like behavior, one will observe only minor changes.

Henceforth, we focus on the vicinity of a cone tip at $\varepsilon=0$ and
$A=A_0$, where the tip position $A_0$ depends on $\Delta$ and $\Omega$.  In
the limit of large driving frequency, the ratio $A_0/\Omega$ matches a zero
of the zeroth-order Bessel function of the first kind.  Here, we do not
restrict ourselves to this limit, such that the possible values of $A_0$
have to be determined numerically.  At the tip, the Floquet states have
zero quasienergy and definite generalized parity.  We denote these states by
$|\varphi_\pm(t)\rangle$ with mean energies $E_\pm$ and
use them states as basis.  Then the Floquet Hamiltonian becomes
\begin{equation}
\mathcal{H}' = \frac{a}{2}\sigma_z' + \frac{b}{2}\sigma_x'
\label{H'}
\end{equation}
with the effective parameters
\begin{align}
a ={}& \frac{A_0-A}{A_0}
	\big(2E_- - \llangle\varphi_-|\sigma_x|\varphi_-\rrangle \big) ,
\label{a}
\\
b = {}& \varepsilon\llangle\varphi_-|\sigma_z|\varphi_+\rrangle ,
\label{b}
\end{align}
where the prime refers to the new time-dependent basis.  Notice that this
constitutes an approximation, because a basis of the full Sambe space
contains also all equivalent Floquet states $e^{-in\Omega
t}|\varphi_\pm(t)\rangle$.  The eigenvalues of $\mathcal{H}'$ are the
quasienergies $\pm r/2$ where $r = \sqrt{a^2+b^2}$.  The main ideas of the
perturbation theory are summarized in the appendix, while details can be
found in Ref.~\onlinecite{KohlerPRA24}.

\subsection{Fermionic environment and master equation}

To describe the open two-level system, we employ the many-particle
version of $H(t)$,
\begin{equation}
H_\text{wire}(t) = \sum_{n,m} h_{nm}(t) c_n^\dagger c_m \,,
\end{equation}
where ``wire'' refers to the central system without the leads. Here, $c_n$
is the fermionic annihilation operator of an electron at site $n$, and
$h_{nm}$ denotes the matrix elements of the single-particle
Hamiltonian~\eqref{H1p}.

The environment consists of leads modeled as Fermi seas with Hamiltonians
$H_\ell = \sum_q \lambda_{\ell q}c_{\ell q}^\dagger c_{\ell q}$ which are
in a grand canonical state with chemical potential $\mu_\ell$, where
$\ell=L,R$ is used to label both the leads and the site coupled to it.
Hence, $\langle c_{\ell q}^\dagger c_{\ell q} \rangle =
f(\epsilon_q-\mu_\ell)$ with the Fermi function $f(x) =
[e^{x/k_BT}+1]^{-1}$.  The coupling between site $|\ell\rangle$ and lead
$\ell$ is established by the tunnel Hamiltonian $V_\ell = \sum_q
\lambda_{\ell q} c_\ell^\dagger c_{\ell q} + \text{h.a.}$  It is fully
specified by the incoherent tunnel rate $\Gamma_\ell(\epsilon) = 2\pi\sum_q
\lambda_{\ell q}^2 \delta(\epsilon-\epsilon_q)$, which we assume energy
independent.\cite{BruusFlensberg2004}

Following Refs.~\onlinecite{LehmannPRL02, LehmannJCP03}, we eliminate the
leads within second-order perturbation theory to obtain a Bloch-Redfield
equation for the reduced density operator of the wire.  To benefit from the
knowledge acquired from Floquet theory, it is convenient to define the
Floquet annihilation operators,
\begin{equation}
c_\alpha(t) = \sum_n \langle\phi_\alpha(t)|n\rangle c_n
\end{equation}
which at equal time obey the usual fermionic anti-commutation relations.
They allows us to define the Floquet single-particle density operator
$R_{\alpha\beta} = \langle c_\beta^\dagger(t) c_\alpha(t)\rangle$ whose
equation of motion can be derived from the Bloch-Redfield equation.

For very weak coupling one can apply a rotating-wave approximation which
ignores off-diagonal elements of $R$ such that $R_{\alpha\beta} = P_\alpha
\delta_{\alpha,\beta}$ with the Floquet state populations $P_\alpha$.
In the long-time limit, the populations become practically time-independent
and obey \cite{LehmannJCP03}
\begin{equation}
\sum_{\ell,k} w^{(\ell)}_{\alpha k} P_\alpha
= \sum_{\ell,k} w^{(\ell)}_{\alpha k}
  f(q_\alpha+k\Omega-\mu_\ell) .
\label{MErwa}
\end{equation}
The transitions rates $w^{(\ell)}_{\alpha k} = \Gamma_\ell
|\langle\ell|\phi_{\alpha k}\rangle|^2$ are governed by the sideband
resolved overlap of the localized wire state $|\ell\rangle$ with the
Floquet state $|\phi_\alpha\rangle$.
This equation for the $P_\alpha$ will be used for analytical
considerations, while all numerical results are computed with the full
master equation for $R_{\alpha\beta}(t)$.

An alternative derivation of Eq.~\eqref{MErwa} starts from a formulation
with Green's functions for which the Floquet equation of the two-level
system contains a self-energy stemming from the coupling to the leads.
\cite{KohlerPR05, ArracheaPRB06, StefanucciPRB08} It allows a treatment
beyond weak site-lead coupling and, thus, considers the resulting level
broadening.  It has been shown that for weak coupling, the
Floquet-Green's function approach becomes equivalent to the present master
equation. \cite{KohlerPR05} Notice that strong coupling is not
considered here, because already without the driving, it causes significant
deviations from a Gibbs state.\cite{ChungPRB01, CheongPRB04, SharmaPRB15}

For the numerical calculations, we first determine the smallest value $A_0$
at which the quasienergies for $\varepsilon=0$ cross, i.e., the position of
the cone tip.  On a circle around the tip with $r\ll\Delta$, the
quasienergies are (approximately) constant.  To determine the corresponding
detuning $\varepsilon$ and amplitude $A$, we parametrize the circle as $a =
r\cos\vartheta$ and $b=r\sin\vartheta$, where the angle $\vartheta=0$
stands for a driving amplitude slightly below $A_0$, while
$\vartheta=\pi/2$ and $3\pi/2$, mean $A=A_0$ and $\varepsilon\neq 0$.
Equations~\eqref{a} and \eqref{b} provide the corresponding values of
$\varepsilon$ and~$A$.

\section{Floquet state population}
\label{sec:results}

Since owing to the rotating-wave approximation, the populations of the
Floquet states in Eq.~\eqref{MErwa} are uncoupled, one readily obtains the
stationary solution
\begin{equation}
P_\alpha = \frac{\sum_{\ell,k} w^{(\ell)}_{\alpha k}
f(q_\alpha+k\Omega-\mu_\ell)}{\sum_{\ell,k} w^{(\ell)}_{\alpha k}} .
\label{Pstat}
\end{equation}
Sufficiently close to the Fermi surface, it can be approximated by using
the Taylor expansion
\begin{equation}
f(x) = \frac{1}{2} + \frac{x}{4k_BT}\quad\text{for $|x|\lesssim k_BT$} ,
\label{fsapprox}
\end{equation}
while outside this window, $f(x)$ practically equals zero or unity
depending on the sign of its argument.

\subsection{Equal coupling to both leads}

Let us start the discussion with a symmetric situation close to the one
with a bosonic bath considered in Ref.~\onlinecite{KohlerPRA24}.  We assume
that both leads have equal chemical potentials, $\mu_L=\mu_R=0$, and
equal tunnel coupling to the respective lead, $\Gamma_L = \Gamma_R =
\Gamma$.  Then we can use the relation $\sum_\ell|\ell\rangle\langle\ell| =
\mathbf{1}$ to obtain $\sum_\ell w_{\alpha k}^{(\ell)} = \Gamma
\langle\phi_{\alpha k}|\phi_{\alpha k}\rangle$ and $\sum_{\ell,k} w_{\alpha
k}^{(\ell)} = \Gamma$ such that Eq.~\eqref{Pstat} becomes
\begin{equation}
P_\alpha = \sum_k \langle\phi_{\alpha k}|\phi_{\alpha k}\rangle
f(q_\alpha+k\Omega) .
\label{Psymm}
\end{equation}
Interestingly, the populations are independent of $\Gamma$, despite that the
wire-lead couplings $V_\ell$ do not commute with the wire Hamiltonian.
This result is consistent with the natural expectation for the undriven
limit in which the occupation probability of the sidebands becomes
$\propto\delta_{k,0}$.  Then $P_\alpha = f(q_\alpha)$ with the quasienergy
$q_\alpha$ being the corresponding eigenenergy of the undriven
Hamiltonian.

Expression \eqref{Psymm} can be simplified in two limits.  First for rather
low temperatures, $k_BT\lesssim\Omega$, such that the Fermi function for all
sideband indices $k<0$ is close unity and practically vanishes for
$k>0$, one may assume that the overlaps $\langle\phi_{\alpha
k}|\phi_{\alpha k}\rangle$ are independent of the sign of $k$ (which is not
exactly fulfilled, but represents a reasonable approximation).  Then
the normalization of the Floquet states yields $\langle\phi_{\alpha
0}|\phi_{\alpha 0}\rangle + 2\sum_{k<0} \langle\phi_{\alpha k}|\phi_{\alpha
k}\rangle = 0$ such that
\begin{equation}
P_\alpha = \frac{1}{2} + \langle\phi_{\alpha 0}|\phi_{\alpha 0}\rangle
\Big(f(q_\alpha)-\frac{1}{2}\Big) ,
\label{lowT}
\end{equation}
i.e., the population is governed by the Fermi function evaluated at the
quasienergy.  However, it becomes $f(q_\alpha)$ only if the Floquet state
has its full weight in the sideband with $k=0$, which is the case only for
$A=0$.  Despite this limitation, we refer to the result in Eq.~\eqref{lowT}
as Floquet-Gibbs limit, because the expression depends only on the
quasienergy in the first Brillouin zone, $q_\alpha$, and not on the
sidebands energies $q_\alpha+k\Omega$ with $k\neq 0$.

For larger temperatures, we assume that for all relevant sidebands,
approximation \eqref{fsapprox} holds.  Then,
\begin{equation}
P_\alpha = \frac{1}{2} + \sum_k \langle\phi_{\alpha k}|\phi_{\alpha k}\rangle
\frac{q_\alpha+k\Omega}{4 k_BT}
= f(E_\alpha)
\label{highT}
\end{equation}
where we have used expression \eqref{idt} for the mean energy.  As we will
see below in our numerical example, Eq.~\eqref{highT} predicts the correct
behavior for intermediate and large temperatures.

\begin{figure}
\centerline{\includegraphics{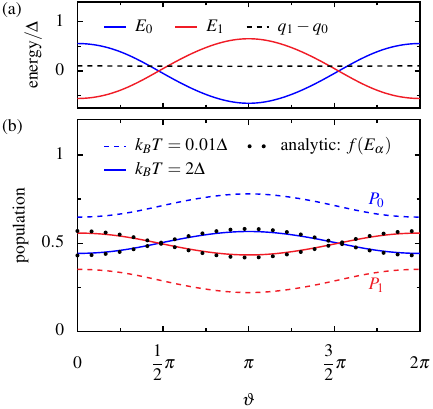}}
\caption{(a) Mean energies and quasienergy splitting on a circle around the
cone tip at the first crossing for driving frequency $\Omega=3\Delta$.  The
angle $\vartheta=0$ corresponds to $\varepsilon=0$ and a driving amplitude
$A<A_0$, i.e., slightly smaller than the one at the cone tip.
(b) Populations for equal coupling to two leads with tunnel rate
$\Gamma_L=\Gamma_R=\Delta/10$ and chemical potential $\mu=0$.}
\label{fig:angle}
\end{figure}

For a numerical confirmation, we focus on the quasienergy crossing shown in
Fig.~\ref{fig:cone}(b) and choose the detuning and the amplitude on a
circle around the tip with $r=0.1\Delta$ as described at the end of
Sec.~\ref{sec:model}.  On this circle, the mean energies shown in
Fig.~\ref{fig:angle}(a) are interchanged as discussed above.  The
corresponding populations for a relatively small driving frequency
$\Omega=3\Delta$ and both low and intermediate temperature are shown in
Fig.~\ref{fig:angle}(b).  One can appreciate that the approximation in
Eq.~\eqref{highT} fits the numerical result rather well, i.e., a
mean-energy state is established.  It is also worth mentioning that the
continuous interchange of the mean energies and the populations implies
that for some value of $\vartheta$, $E_0=E_1$ and $P_0 = P_1 = 1/2$.  The
latter means that the density operator has maximal entropy even at
intermediate temperatures.  In the absence of driving, such behavior is
expected only when the thermal energy by far exceeds all energy splittings.
For the driven two-level system in the vicinity of conical intersections,
it has been found also for dissipation stemming from an Ohmic heat
bath.\cite{KohlerPRA24}

In the low-temperature limit $k_BT=0.01\Delta$, owing to the practically
constant quasienergies, Eq.~\eqref{lowT} may lead to the conclusion that
the populations should be constant on the circle.  By contrast, we witness
a clear dependence on $\vartheta$, i.e., the Floquet-Gibbs limit is not
fulfilled.  Also the $\vartheta$-dependence of the zeroth sideband,
$\langle\phi_{\alpha 0}|\phi_{\alpha 0}\rangle$, cannot explain the
discrepancy.

\begin{figure}
\centerline{\includegraphics{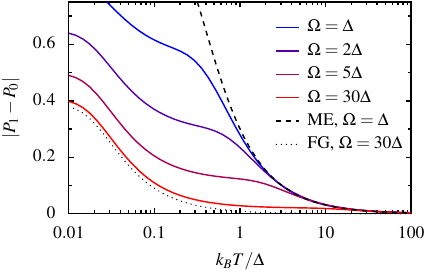}}
\caption{Population difference as a function of the temperature for
detuning $\varepsilon=0$, amplitude $A>A_0$ (i.e., $\vartheta=\pi$), and
various driving frequencies.  The dashed and the dotted lines mark the
conjectures of a Floquet-Gibbs limit (FG) and a mean-energy state (ME) in
Eqs.~\eqref{lowT} and \eqref{highT}, respectively.  The mean energy is
practically the same for all driving frequencies considered.  All other
parameters are as in Fig.~\ref{fig:angle}.}
\label{fig:overlap}
\end{figure}

To demonstrate that this discrepancy can be attributed to the relatively
small driving frequency, we compare the conjectures \eqref{lowT} and
\eqref{highT} with the numerical results for different driving frequencies
and temperatures.  As quantity of interest, we consider the population
difference $\Delta P \equiv |P_1-P_0|$ for an amplitude slightly above the
cone tip, i.e., for $\vartheta=\pi$.  The results expected from
Eqs.~\eqref{lowT} and \eqref{highT} are $\Delta P = \langle\phi_{\alpha
0}|\phi_{\alpha 0}\rangle \coth(|q_0|/2k_BT)$ and $\coth(|E_0|/2k_BT)$,
respectively, where we have used $q_0+q_1 = 0 = E_0+E_1$.
Figure~\ref{fig:overlap} shows that the approximation in Eq.~\eqref{highT}
is fulfilled rather well for $k_BT\gtrsim\Omega$.  However, the
Floquet-Gibbs limit (dotted line) only emerges for a relatively large
driving frequency $\Omega\ggg\Delta$.  In this respect, the present
situation is analogous to the one with a bosonic heat
bath.\cite{KohlerPRA24}
The same conclusions can be drawn from the data for $\vartheta=0$.  For
this value, however, quasienergies and mean energies have opposite order,
$q_0<q_1$ while $E_0>E_1$.  This leads to a non-monotonic behavior as a
function of $k_BT$ which is less convenient to analyze.

\subsection{Asymmetric wire-lead couplings}

\begin{figure}
\centerline{\includegraphics{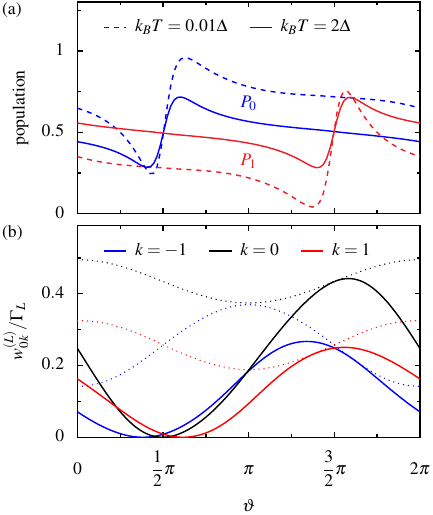}}
\caption{(a) The same as Fig.~\ref{fig:angle}(b) but in the absence of the
right lead, $\Gamma_R=0$, which yields significant deviations from the
conjectures in Eqs.~\eqref{lowT} and \eqref{highT}.
(b) Coefficient $w_{\alpha k}^{(\ell)} = \Gamma_\ell |\langle\ell|\phi_{\alpha
k}\rangle|^2$ appearing in Eq.~\eqref{Pstat} for the Floquet ``ground
state'' $|\phi_0(t)\rangle$ and dot state $|L\rangle$.  The dashed lines
show for comparison the corresponding coefficient for two leads with equal
tunnel rates, $\Gamma_L=\Gamma_R$.}
\label{fig:onelead}
\end{figure}

To highlight the role of equal coupling to both leads, we consider the
asymmetric case with the right lead disconnected, $\Gamma_R=0$.  All other
parameters are as before, such that both the quasienergies and the mean
energies are unchanged.  Nevertheless, the resulting Floquet state
populations shown in Fig.~\ref{fig:onelead}(a) differ qualitatively from
the ones in Fig.~\ref{fig:angle}(b).  Even the reflection symmetry at
$\vartheta=\pi$ is lost.  Obviously, we no longer obtain a
Floquet-Gibbs-like state or a mean energy state.

The impact of the asymmetry can be understood from the analytic result
\eqref{Pstat}.  For a coupling to only the left lead, the projectors
$|\ell\rangle\langle \ell|$ in the coefficients $w_{\alpha k}^{(\ell)}$ no
longer sum up to unity, such that
\begin{equation}
P_\alpha \propto \sum_k |\langle L|\phi_{\alpha k}\rangle|^2
f(q_\alpha+k\Omega).
\label{Pasym}
\end{equation}
Figure ~\ref{fig:onelead}(b) depicts one of the prefactors of the Fermi
function in Eq.~\eqref{Pasym} for the sideband indices $k=0,\pm 1$.  It can
be appreciated that close to $\vartheta=\pi/2$, they are rather small and for
some value of $\vartheta$ practically vanish.  However, they vanish at
different angles.  Therefore, their relative difference may be huge, and
the symmetry with respect to the sign of $k$ underlying the derivation of
Eq.~\eqref{lowT} may be significantly broken.  As a consequence, a
dependence of $P_\alpha$ on some sideband energies will remain.

For the high-temperature approximation which leads to the mean-energy state
the situation is even clearer.  The sideband occupation in the middle of
Eq.~\eqref{highT} now contains the projector $|L\rangle\langle L|$.
Therefore, the summation no longer yields the mean energy.  This is already
gradually the case for $\Gamma_L\neq\Gamma_R$.  This reasoning reveals that
the symmetry of the dot-lead couplings is a crucial ingredient for a
stationary state that can be written as simple function of the quasienergy
or the mean energy.

Let us finally remark that the (approximate) localization of the Floquet
states for $A=A_0$ and small but non-vanishing detuning $\epsilon$ (i.e.,
$\vartheta=\pi/2$ and $3\pi/2$) that explains the data in
Fig.~\ref{fig:onelead}(b) can be understood form the perturbation theory in
Sec.~\ref{sec:cone}.  Since at the tip, the zeroth order of the Floquet
Hamiltonian vanishes, the Floquet states are essentially determined by the
detuning, which is proportional to $\sigma_z$.  Hence the natural
expectation is a tendency towards localized states.

\section{Conclusions}
\label{sec:conclusions}

We have addressed the question whether for a driven quantum system coupled
to electron reservoirs, Floquet-Gibbs states or mean-energy states emerge.
As a basic model, we have employed a two-level system that can be
occupied with up to two spinless electrons without taking their interaction
into account.  This situation allows a good comparison with a former study
\cite{KohlerPRA24} in which dissipation stems from a coupling to a bosonic
heat bath.  Also here we have focussed on conical intersections, because in
their vicinity, quasienergies and mean energies behave rather differently
and in a characteristic manner.  Then much insight can be gained from the
Bloch-Redfield master equation for the single-particle density operator.
Within a rotating-wave approximation it has been solved analytically.

A main observation is that no true Floquet-Gibbs states are found, i.e.,
the populations match the Fermi function evaluated at the quasienergies
only in trivial limits such as in the absence of driving.  Nevertheless,
there are situations in which the deviation is determined by the
quasienergy in the first Brillouin zone, such that one may speak of
Floquet-Gibbs-like states.  As for the dissipative two-level system, this
requires rather large driving frequencies.  For smaller frequencies and
intermediate temperatures, the stationary populations are directly given by
a Fermi function with the mean energy of the Floquet states as argument.
Thus, also here the mean-energy states seem rather generic, at least under
the condition of a sufficiently symmetric setup.  In particular, the tunnel
couplings of the sites to the respective leads have to be approximately equal.
For larger systems, for example for models that describe conducting
molecules, such equal coupling of all sites to an electron reservoir may be
essential for the emergence of Floquet-Gibbs-like states or mean-energy
states.

Our prediction of certain reduced density operators may be tested in
any experiment that is sensitive to the occupation of Floquet states.  One
may for example proceed as in Ref.~\onlinecite{ChenPRB21} where a driven
closed double quantum dot occupied with a single electron was coupled to a
superconducting cavity.  Then the transmission of the latter provides
information about the occupation of Floquet states.  A corresponding
experiment with an open double dot may provide novel insight to the
stationary state of driven quantum systems.

\begin{acknowledgments}
This work was supported by the Spanish Ministry of Science, Innovation, and
Universities (Grant No.\ PID2023-149072NB-I00), and by the CSIC Research
Platform on Quantum Technologies PTI-001.
\end{acknowledgments}

\section*{Data Availability Statement}
The numerical data shown in the figures is available from the author upon
reasonable request.

\appendix

\section{Floquet Hamiltonian near the cone tip}

For a perturbation theory in the vicinity of the cone tip, we follow
Ref.~\onlinecite{KohlerPRA24} and separate the Floquet Hamiltonian into two
parts, namely the zeroth-order contribution
\begin{equation}
\mathcal{H}_0 = \frac{\Delta}{2} \sigma_x
      + \frac{A_0}{2}\sigma_z \cos(\Omega t) -i\frac{\partial}{\partial t}
\equiv
      H_0(t) -i\frac{\partial}{\partial t},
\end{equation}
with $A_0$ such that the quasienergies cross, i.e., they are degenerate and
vanish, $q_0 = q_1 = 0$.  Since $\mathcal{H}_0$ obeys generalized parity,
$G\mathcal{H}_0G^{-1} = \mathcal{H}_0$, its two non-equivalent Floquet
states can be classified as even or odd.  They will be denoted by
$|\varphi_\pm(t)\rangle$ and their mean energies bay $E_\pm$.  The perturbation
\begin{equation}
\mathcal{H}_1 = \frac{\varepsilon}{2}\sigma_z
+ \frac{A-A_0}{2}\sigma_z \cos(\Omega t)
\label{app:H1}
\end{equation}
will be treated with degenerate perturbation theory.  To do so, we
calculate the matrix elements of $\mathcal{H}_1$.  Since $G\sigma_z G^{-1}
= -\sigma_z$ and $G\sigma_z\cos(\Omega t)G^{-1} = \sigma_z\cos(\Omega t)$,
in the new basis the diagonal matrix elements of the first term in
Eq.~\eqref{app:H1} and the off-diagonal elements of the second term vanish.

The remaining matrix elements of $\sigma_z$ can be evaluated
straightforwardly to yield $b/2$ with $b$ given in Eq.~\eqref{b}.  The
diagonal elements of the time-dependent term can be obtain upon noticing
that
\begin{equation}
\sigma_z\cos(\Omega t) = \big(2H_0(t) - \Delta\sigma_x \big)/A_0 .
\end{equation}
Together with the relation $E_- =
\llangle\varphi_-|H_0(t)|\varphi_-\rrangle = -E_+$ follows the first term
of the effective Floquet Hamiltonian \eqref{H'} with the coefficient $a$ in
Eq.~\eqref{a}.

\end{document}